\documentclass[10pt,twocolumn,journal,final]{IEEEtran}
\usepackage{amsfonts,cite}
\usepackage{caption2}

\usepackage{stfloats}
\usepackage{color}

\usepackage{algorithm}
\usepackage{algorithmic}
\usepackage{multirow}
\usepackage[caption=false,font=footnotesize]{subfig}
\usepackage[cmex10]{amsmath}
\usepackage{cite}

\ifCLASSINFOpdf

\else

   \usepackage[dvips]{graphicx}

   \graphicspath{{../fig/}}

   \DeclareGraphicsExtensions{.eps}
\fi

\usepackage{array}
\usepackage{mdwmath}
\usepackage{mdwtab}

\usepackage{url}

\begin{document}

\title{A Simple Two-stage Equalizer With Simplified Orthogonal Time Frequency Space Modulation Over Rapidly Time-varying Channels}

\author{
Li~(Alex)~Li,
       Hua Wei,
       Yao Huang,
       Yao Yao,
       Weiwei Ling,
       Gong Chen,
       Peng Li,and Yunlong Cai

}

\maketitle

\begin{abstract}
In this work, we derive a equivalent delay-Doppler channel matrix of the Orthogonal Time Frequency Space (OTFS) modulation that has not been studied in previous literature. It has the similar structure as the banded channel matrix of OFDM systems over rapidly time-varying channels. However, the band in the equivalent channel matrix will no longer spread with the increase of the Doppler spread once the length of maximum channel delay spread and the OTFS frame duration are determined. Furthermore, the equivalent channel matrix can simplify the OTFS modulation in the transmitter side. Incorporating the equivalent channel matrix, we propose a simple two-stage equalizer in $1$ dimensional operations for OTFS modulation. First, the receive signal is equalized using the conventional OFDM single-tap equalizer in the frequency domain. The multipath effects can be removed. In the second stage, another low complexity delay-Doppler domain equalizer is employed to eliminate the effects of the residual interference caused by the Doppler spread with the equivalent channel matrix. The simulation results demonstrate that the proposed method is superior to the conventional single-tap equalizer and full minimum mean squared error (MMSE) equalizer of OFDM systems in terms of BER in high Doppler spread scenarios.
\end{abstract}


\section{Introduction}

\IEEEPARstart{W}{ith} the increasing demand of mobile communications service, the requirement of the systems with high spectral efficiency in high Doppler spread scenarios must be met. There are several alternatives to the conventional orthogonal frequency division multiplexing (OFDM) scheme proposed. Most modulation schemes can effectively reduce the out of band leakage and increase the spectral efficiency compared to the conventional OFDM scheme, but are still vulnerable to the high Doppler spread as shown in \cite{cai2017modulation}. In contrast to the schemes stated above, Orthogonal Time Frequency Space (OTFS) modulation is one of promising techniques, which is more robust to high Doppler spread and phase noise and provides high diversity order \cite{hadani2017orthogonal}. However, only an overall framework of designing OTFS has been given. The exact implementation is not very clear compared to OFDM systems as stated in \cite{deannew}, because the effects of the delay spread and Doppler spread on the equivalent channel matrix is not studied.

In this paper, we aim to derive a equivalent channel matrix of OTFS modulation in a general form, which can help us design low complexity transmitter and receiver even in severe rapidly time-varying channels, e.g., high-speed railway mobile communications. In this case, the conventional single-tap equalizer for OFDM systems is not working. We propose a communication system equipped with one transmit antenna and one receive antenna. The cyclic-prefix (CP) is added at the start of each transmit OFDM symbol to ensure the multipath effects eliminated using single-tap equalizer in the frequency domain. Besides, the effects of time variations can be mitigated using interference cancellation scheme with a equivalent channel matrix in a simple form.
\begin{table}[!htbp]
 \begin{center}
     \begin{tabular}{| l | l | }
     \hline
     Notation & Description \\
     \hline
     $\mathbf{A}\vert_{N \times M}$  & a $N \times M$ matrix \\
     \hline
     $\mathbf{a}\vert_{N \times 1}$  & a $N \times 1$ vector   \\
     \hline
      $\mathbf{I}_{N}$& a $N \times N$ identity matrix  \\
     \hline
     $(\cdot)^{H}$& Hermitian transpose of a matrix or a vector\\
     \hline
    $(\cdot)^{T}$& transpose of a matrix or a vector\\
     \hline
     $(\cdot)^{\ast}$ & complex conjugate\\
     \hline
     $\lfloor \cdot \rfloor$& flooring operation\\
     \hline
     $\otimes$& Kronecker product\\
     \hline
     \end{tabular}
 \end{center}
 \caption{Mathematical Notation}
 \label{Tab:Notation}
 \end{table}

\section{System Model}
\subsection{Transmitter Processing}
Basically, the system model is very similar to the convectional OFDM system. For brevity, we omit the detailed description of OFDM, and represent the system model in the matrix form. First, the transmission symbols $x(\nu,l)$ are spread over the delay-Doppler domain, and the symbols is then converted to the time-frequency symbols $x(k,n)$ via discrete symplectic Fourier transform (DSFT). The DSFT can be equivalent to the operations of inverse fast Fourier transform (IFFT) $ \mathbf{F}^{H}\vert_{N_{\nu}\times N_{\nu}}$ plus fast Fourier transform (FFT) $\mathbf{F}\vert_{N_{l}\times N_{l}}$. Hence, the time-frequency symbol matrix $\mathbf{X}_{kn}\vert_{N_{l} \times N_{\nu}}$ can be represented as
\begin{equation}\label{Eqn:DSFT_mat}
\mathbf{X}_{kn}\vert_{N_{l} \times N_{\nu}} = \mathbf{F}\vert_{N_{l}\times N_{l}}(\mathbf{F}^{H}\vert_{N_{\nu}\times N_{\nu}} \mathbf{X}_{\nu l}\vert_{N_{\nu} \times N_{l}})^{T},
\end{equation}
where $\mathbf{X}_{\nu l}\vert_{N_{\nu} \times N_{l}}$ denotes the delay-Doppler symbol matrix with the size of $N_{\nu}\times N_{l} $. After the IFFT of $\mathbf{X}_{kn}\vert_{N_{l} \times N_{\nu}} $, the signal matrix $\mathbf{X}_{t}\vert_{N_{l} \times N_{\nu}}$ in the time domain can be obtained. The CP is added at the beginning of each column vector in the matrix $\mathbf{X}_{t}\vert_{N_{l} \times N_{\nu}}$, and then the matrix with the CP is vectorized into one column vector for the multipath channels. For the receiver side, the inverse operations will be performed after the removal of the CP in the matrix form.  The above operations can be efficiently performed. However, it is not easy to derive the equivalent delay Doppler channel matrix. In order to derive the equivalent matrix for the simple equalizer design , the operations in (\ref{Eqn:DSFT_mat}) needs to be vectorized column by column as follows:
\begin{equation}\label{Eqn:DSFT_vec}
\mathbf{x}_{kn}\vert_{N_{l}N_{\nu} \times 1} = \mathbf{\bar{F}}\vert_{N_{l}N_{\nu}\times N_{l}N_{\nu}} (\mathbf{I}_{N_{l}} \otimes \mathbf{F}^{H}\vert_{N_{\nu}\times N_{\nu}}) \mathbf{x}_{\nu l}\vert_{N_{\nu}N_{l}\times 1},
\end{equation}
where $\mathbf{x}_{kn}\vert_{N_{l}N_{\nu} \times 1}$ denotes the time-frequency signal vector that has $N_{\nu}$ OFDM symbols, and each OFDM symbol consist of $N_{l}$ subcarriers, $\mathbf{\bar{F}}\vert_{N_{l}N_{\nu}\times N_{l}N_{\nu}}$ denotes the extended FFT matrix, and the elements of $\mathbf{F}\vert_{N_{l}\times N_{l}}$ are uniformly distributed in the extended matrix. The $i$th row vector of the matrix $\mathbf{\bar{F}}\vert_{N_{l}N_{\nu}\times N_{l}N_{\nu}}$ is given as
\begin{equation}\label{Eqn:FFT_ext}
\mathbf{\bar{f}}\vert_{1\times N_{l}N_{\nu}}(i) = [\underbrace{\mathbf{0}}_{i-1}, W_{i_{r}0},\underbrace{\mathbf{0}}_{N_{\nu}},W_{i_{r}1},\ldots,W_{i_{r}N_{l}},\underbrace{\mathbf{0}}_{N_{\nu}-1}],
\end{equation}
where $i=0,1,\ldots,N_{l}-1$ and $i_{r} =\lfloor \frac{i}{N_{v}}\rfloor$ represents the integer quotient of $i$ by $N_{v}$, and $W_{kl}=e^{-j2\pi\frac{kl}{N_{l}}}$. After the matrix operation in (\ref{Eqn:DSFT_vec}), the time-frequency symbol vector has been obtained and can be converted to the signal vector in the time domain via the extended IFFT $\mathbf{\bar{F}}^{H}\vert_{N_{l}N_{\nu}\times N_{l}N_{\nu}}$ as follows:
\begin{equation}\label{Eqn:IFFT_vec}
\mathbf{x}_{t}\vert_{N_{l}N_{\nu} \times 1} = \mathbf{\bar{F}}^{H}\vert_{N_{l}N_{\nu}\times N_{l}N_{\nu}}\mathbf{x}_{kn}\vert_{N_{l}N_{\nu} \times 1}.
\end{equation}
However, the vector $\mathbf{x}_{t}\vert_{N_{l}N_{\nu} \times 1}$ is not in a sequential manner in the time domain due to the matrix operations described above. There is an additional reordering matrix required. The elements in the reordering matrix $\boldsymbol{\Xi}\vert_{N_{l}N_{\nu} \times N_{l}N_{\nu}}$ is given as
\begin{equation}\label{Eqn:reoder_mat}
\Xi(i,j) =
\begin{cases}
&1~~\textrm{if}~j=\lfloor\frac{i}{N_{l}}\rfloor+\textrm{mod}(i,N_{l})\cdot N_{v} \\
&0~~\textrm{otherwise}
\end{cases} ,
\end{equation}
where $i,j = 0,1,\ldots,N_{v}N_{l}-1$, and the symbol $\textrm{mod}(\cdot)$ denotes the modular operation. So the vector $\mathbf{x}_{t}\vert_{N_{l}N_{\nu} \times 1}$ is reordered by the matrix multiplication $\mathbf{\tilde{x}}_{t}\vert_{N_{l}N_{\nu} \times 1} = \boldsymbol{\Xi}\vert_{N_{l}N_{\nu} \times N_{l}N_{\nu}}\mathbf{x}_{t}\vert_{N_{l}N_{\nu} \times 1}$. Note that the CP will be added at every $N_{v}$ element.

Similar to the time-varying multipath channel model given in \cite{schniter2004low,rugini2005simple}, we only consider the channel impulse response (CIR) after the removal of the CP in the sense that the size of the channel matrix reduces to $N_{l}N_{\nu} \times N_{l}N_{\nu}$. The receive signals in the time domain can be expressed as
\begin{equation}\label{Eqn:recv_sig_t}
\mathbf{y}_{t}\vert_{N_{l}N_{\nu} \times 1} = \mathbf{H}_{tl}\vert_{N_{l}N_{\nu} \times N_{l}N_{\nu} }  \mathbf{\tilde{x}}_{t}\vert_{N_{l}N_{\nu} \times 1}+\mathbf{n}\vert_{N_{l}N_{\nu} \times 1},
\end{equation}
where the matrix $\mathbf{H}_{tl}\vert_{N_{l}N_{\nu} \times N_{l}N_{\nu} } $ represents the time-varying amultipath channel model, the element of which $h_{i,j}$ is the channel impulse response at the $i$th time interval and the $j$th path, and $i=0,1,\ldots,N_{v}N_{l}-1$, $j= i-L+1,j-L+2,\ldots,i$. If $j<0$, $j=j+N_{\nu}N_{l}$. The length of the maximum delay spread of the channel is $L$.  The elements in the vector $\mathbf{n}\vert_{N_{l}N_{\nu} \times 1}$ represent the additive white Gaussian noise (AWGN) vector with zero mean value and $\sigma_{n}^{2}$ variance.

\subsection{Receiver Processing}
In (\ref{Eqn:recv_sig_t}), the receive signals in the time domain are obtained in a sequential manner, which can not be directly processed by the following blocks of the receiver. We therefore reorder the the receive signal vector back with the transpose of the reordering matrix $ \boldsymbol{\Xi}\vert_{N_{l}N_{\nu} \times N_{l}N_{\nu}}$. In the next, the inverse operations of the transmitter will be performed. The time-frequency signal vector $\mathbf{y}_{kn}\vert_{N_{l}N_{\nu} \times 1}$ is obtained by the extended FFT $\mathbf{\bar{F}}\vert_{N_{l}N_{\nu}\times N_{l}N_{\nu}}$:
\begin{equation}\label{Eqn:FFT_ext}
\mathbf{y}_{kn}\vert_{N_{l}N_{\nu} \times 1} = \mathbf{\bar{F}}\vert_{N_{l}N_{\nu}\times N_{l}N_{\nu}}\boldsymbol{\Xi}^{T}\vert_{N_{l}N_{\nu} \times N_{l}N_{\nu}} \mathbf{y}_{t}\vert_{N_{l}N_{\nu} \times 1},
\end{equation}
and the inverse operation of (\ref{Eqn:DSFT_vec}) is expressed as
\begin{equation}\label{Eqn:IDSFT_vec}
\mathbf{y}_{\nu l}\vert_{N_{l}N_{\nu} \times 1} =(\mathbf{I}_{N_{l}} \otimes \mathbf{F}\vert_{N_{\nu}\times N_{\nu}}) \mathbf{\bar{F}}^{H}\vert_{N_{l}N_{\nu}\times N_{l}N_{\nu}}\mathbf{y}_{kn}\vert_{N_{l}N_{\nu} \times 1}.
\end{equation}
\subsection{The Equivalent Channel Matrix}
\begin{figure}[!htb]
\captionstyle{center}
\centering
\includegraphics[width=0.5\textwidth]{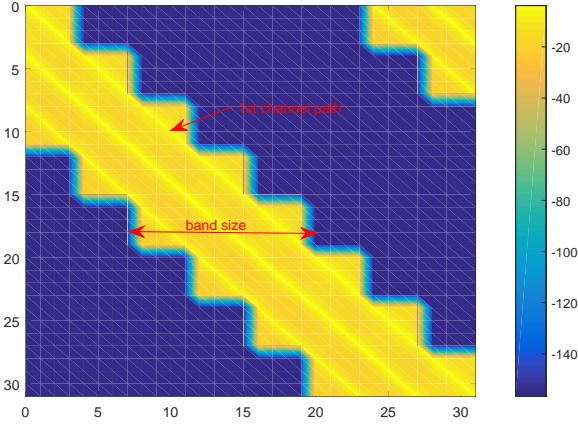}
\caption{The equivalent channel matrix $\mathbf{H}_{eq}$ in the delay Doppler domain with $N_{l}=8, N_{\nu} = 4$, $L=3$ Doppler frequency $f_{d} = 6$KHz}
\label{fig:ch_eq}
\end{figure}

From (\ref{Eqn:DSFT_vec}) to (\ref{Eqn:IDSFT_vec}), all operations are performed by the matrix multiplications. The equivalent channel matrix for OTFS modulation can be given by
\begin{equation}\label{Eqn:H_eq}
\begin{aligned}
\mathbf{H}_{eq}\vert_{N_{l}N_{\nu}\times N_{l}N_{\nu}}
&= \underbrace{(\mathbf{I}_{N_{l}} \otimes \mathbf{F}\vert_{N_{\nu}\times N_{\nu}}) \mathbf{\bar{F}}^{H}\vert_{N_{l}N_{\nu}\times N_{l}N_{\nu}}}_{\mathbf{P}_{1}}\\
&\underbrace{\mathbf{\bar{F}}\vert_{N_{l}N_{\nu}\times N_{l}N_{\nu}}\boldsymbol{\Xi}^{T}\vert_{N_{l}N_{\nu} \times N_{l}N_{\nu}}}_{\mathbf{P}_{0}}\\
&\mathbf{H}_{tl}\vert_{N_{l}N_{\nu} \times N_{l}N_{\nu} }\\
&\underbrace{\boldsymbol{\Xi}\vert_{N_{l}N_{\nu} \times N_{l}N_{\nu}}\mathbf{\bar{F}}^{H}\vert_{N_{l}N_{\nu}\times N_{l}N_{\nu}}}_{\mathbf{Q}_{0}}\\
&\underbrace{\mathbf{\bar{F}}\vert_{N_{l}N_{\nu}\times N_{l}N_{\nu}} (\mathbf{I}_{N_{l}} \otimes \mathbf{F}^{H}\vert_{N_{\nu}\times N_{\nu}})}_{\mathbf{Q}_{1}}. \\
 \end{aligned}
\end{equation}
Due to $\mathbf{\bar{F}}^{H}\vert_{N_{l}N_{\nu}\times N_{l}N_{\nu}}\mathbf{\bar{F}}\vert_{N_{l}N_{\nu}\times N_{l}N_{\nu}}=\mathbf{I}_{N_{l}N_{\nu}}$, the matrix multiplication in (\ref{Eqn:H_eq}) can be simplified as
\begin{equation}\label{Eqn:H_eq_sim}
\begin{aligned}
\mathbf{H}_{eq}\vert_{N_{l}N_{\nu}\times N_{l}N_{\nu}}
&= \mathbf{I}_{N_{l}} \otimes \mathbf{F}\vert_{N_{\nu}\times N_{\nu}} \\
&\boldsymbol{\Xi}^{T}\vert_{N_{l}N_{\nu} \times N_{l}N_{\nu}}
\mathbf{H}_{tl}\vert_{N_{l}N_{\nu} \times N_{l}N_{\nu} }
\boldsymbol{\Xi}\vert_{N_{l}N_{\nu} \times N_{l}N_{\nu}}\\
& \mathbf{I}_{N_{l}} \otimes \mathbf{F}^{H}\vert_{N_{\nu}\times N_{\nu}}.\\
 \end{aligned}
\end{equation}
From (\ref{Eqn:H_eq_sim}), we can observe that the OTFS modulation can be extremely simplified with the reordering matrix and multiple IFFT and FFT operations for the entire OTFS frame duration. Hence, the system model with the equivalent channel matrix in (\ref{Eqn:H_eq_sim}) can be reduced to
\begin{equation}\label{Eqn:sys_mdl_sim}
\mathbf{y}_{\nu l}\vert_{N_{\nu}N_{l} \times 1}=\mathbf{H}_{eq}\vert_{N_{l}N_{\nu}\times N_{l}N_{\nu}}\mathbf{x}_{\nu l}\vert_{N_{\nu}N_{l}\times 1}+\mathbf{n}_{\nu l}\vert_{N_{\nu} N_{l}\times 1},
\end{equation}
where the vector $\mathbf{n}_{\nu l}\vert_{N_{l}N_{\nu} \times 1}$ is the AWGN vector in the delay-Doppler domain. The effects of the channel Doppler spread and delay spread can be seen in the equivalent matrix plotted in Fig. \ref{fig:ch_eq}. The figure is plotted based on $\vert h(\nu,l)\vert$ in the delay-Doppler domain. Although the Doppler spread and delay spread do exist in the channel matrix, the special channel structure does allow a simple equalizer. The band size of the equivalent channel matrix is determined by the parameter $N_{v}(L+1)$. Additionally, it can be observed that the OTFS modulation in the transmitter side can be simplified significantly with the reordering matrix $\boldsymbol{\Xi}\vert_{N_{l}N_{\nu} \times N_{l}N_{\nu}}$ and multiple IFFT. Compared to (\ref{Eqn:DSFT_vec}), the DSFT and IFFT are not required in (\ref{Eqn:H_eq_sim}).

\section{Equalizers for OTFS modulation}
In this part, a simple two-stage equalizers is proposed to eliminate the delay spread and Doppler spread effects. In the first stage, the convectional single-tap equalizer is employed in the frequency domain to remove the multipath effects, and then we use another equalizer to cancel residual interference in the delay-Doppler domain after the first stage equalization.
\subsection{Frequency Domain Equalizer}
As discussed above, the frequency domain equalizer (FDE) is similar to the convectional OFDM single-tap equalizer, which can be represented as
\begin{equation}\label{Eqn:eq_fd}
G_{0}(k,n)=\dfrac{H_{0}^{\ast}(k,n)}{\vert H_{0}(k,n) \vert + \gamma_{\textrm{FD}} },
\end{equation}
where $H^{\ast}(k,n),k=0,1,\ldots,N_{l}-1,n=0,1,\ldots,N_{\nu}-1$ represents complex conjugate of the channel frequency response (CFR) at the $k$th subcarrier and the $n$th OFDM symbol. Note that the IFFT and FFT are employed on the top of DSFT, and the receive signal vector  $\mathbf{y}_{k n}\vert_{N_{l}N_{\nu} \times 1}$  consists of $N_{\nu}$ OFDM symbols. The regularization parameter $\gamma_{\textrm{FD}}$ is similarly defined as \cite{zakharov2015ofdm} plus the noise variance $\sigma^{2}_{n}$.  In the diagonal matrix form, the CFR can be derived as
\begin{equation}\label{Eqn:eq_fd_mat}
\mathbf{H}_{0}(n)\vert_{N_{l}\times N_{l}}=\mathcal{D}( \mathbf{F}\vert_{N_{l}\times N_{l}}\mathbf{H}_{tl}(n)\vert_{N_{l} \times N_{l} }\mathbf{F}^{H}\vert_{N_{l}\times N_{l}}),
\end{equation}
where the symbol $\mathcal{D}(\cdot)$ denotes the diagonal matrix constructed by the diagonal elements of the matrix in the parentheses. Hence, the coefficients of the equalizer $\mathbf{G}_{0}$ can be derived according to (\ref{Eqn:eq_fd}) and (\ref{Eqn:eq_fd_mat}) for different OFDM symbols. Note that we only consider the diagonal elements of the matrix $\mathbf{H}_{0}\vert_{N_{l}\times N_{l}}$. Hence, the above operations are equivalent to the implementation of multiple FFT and IFFT. The receive signal vector after the frequency domain equalizer can be expressed as
\begin{equation}\label{Eqn:recv_sig_eq_fd}
\mathbf{\tilde{y}}_{kn}\vert_{N_{l}N_{\nu} \times 1} = \mathbf{G}_{0} \mathbf{y}_{kn}\vert_{N_{l}N_{\nu} \times 1}.
\end{equation}
Incorporating the same processing in (\ref{Eqn:IDSFT_vec}), the equalized receive signal vector in (\ref{Eqn:recv_sig_eq_fd}) after DSFT is expressed as
\begin{equation}\label{Eqn:recv_sig_eq_fd_IDSFT}
\mathbf{\tilde{y}}_{\nu l}\vert_{N_{l}N_{\nu} \times 1} = \mathbf{P}_{1}\mathbf{\tilde{y}}_{kn}\vert_{N_{l}N_{\nu} \times 1}.
\end{equation}

\subsection{Delay-Doppler Domain Equalizer}

Based on the derived channel matrix in (\ref{Eqn:H_eq_sim}), another delay Doppler domain equalizer (DDE) can be designed to further remove the additional interference from the neighbour symbols in the delay-Doppler domain. With the autocorrelation matrix  $\mathbf{R}_{HH} =\mathbf{H}^{H}_{eq}\vert_{N_{l}N_{\nu}\times N_{l}N_{\nu}}\mathbf{H}_{eq}\vert_{N_{l}N_{\nu}\times N_{l}N_{\nu}}$ Omitting the diagonal elements and clipping other elements with low power to zeros, i.e., $\mathbf{\bar{R}}_{HH}$ , the delay-Doppler equalizer can be expressed as
\begin{equation}\label{Eqn:eq_dd}
\mathbf{\hat{x}}_{\nu l}\vert_{N_{\nu} \times N_{l}}=\mathbf{H}^{H}_{eq}\vert_{N_{l}N_{\nu}\times N_{l}N_{\nu}}\mathbf{y}_{\nu l}\vert_{N_{l}N_{\nu} \times 1}-\mathbf{\bar{R}}_{HH} \lceil\mathbf{\tilde{y}}_{\nu l}\vert_{N_{l}N_{\nu} \times 1}\rfloor,
\end{equation}
where the notation $\lceil \cdot \rfloor$ denotes the quantization operation that maps the receive signal to the nearest constellation points.

\subsection{Remark}
In the above subsections, we propose a two-stage equalizer to mitigate the effects of the delay spread and Doppler spread based on the CFR and the equivalent channel matrix in (\ref{Eqn:H_eq_sim}). In contrast to the methods given in \cite{hadani2017orthogonal}, the coarse estimates of the transmit symbols are obtained by the single-tap equalizer in the frequency domain. Additionally, the coefficients of the proposed single-tap equalizer are computed by FFT and IFFT for each OFDM symbol. Its complexity is much lower than $2$ dimensional periodic convolution of the windowed channel response. For the fine estimates of the transmit symbols, we employ the second stage equalizer, which perform the interference cancellation using $1$ dimensional strategy. Although the interference cancellation scheme in our work is equivalent to the $2$ dimensional decision feedback equalizer (DFE) used in the previous work. However, it does allow a simpler and straightforward implementation to reconstruct the channel matrix in the delay-Doppler domain.

\section{Simulation Results}
\begin{figure}[!htb]
\captionstyle{center}
\centering
\includegraphics[width=0.5\textwidth]{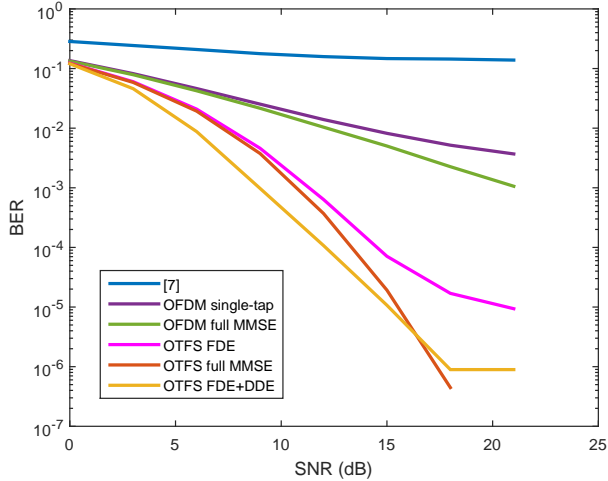}
\caption{BER v.s. SNR with TU$6$ channel model, Doppler frequency $f_{d}=6$KHz }
\label{fig:ber_snr}
\end{figure}

\begin{figure}[!htb]
\captionstyle{center}
\centering
\includegraphics[width=0.5\textwidth]{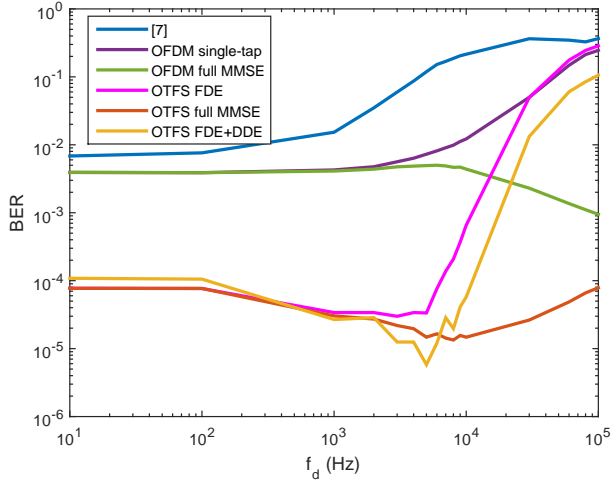}
\caption{BER v.s. Doppler frequency with TU$6$ channel model, SNR$=20$dB}
\label{fig:ber_fd}
\end{figure}

In this section, we evaluate the BER performance of the proposed equalizers (FDE,FDE+DDE), full minimum mean squared error (MMSE) equalizer using the equivalent channel matrix of OTFS modulation, and the conventional single-tap equalizers and full MMSE equalizer for OFDM systems. Additionally, we modified the DFE equalizer for static multipath channels proposed in \cite{wei2017universal} for comparison. The simulation parameters can be found in Tab. \ref{Tab:Sim}. BER is averaged by $5000$ simulation trials, and each trial consists of $8192$ QPSK symbols. In Fig. \ref{fig:ber_snr} and \ref{fig:ber_fd}, the proposed FDE and FDE+DDE perform better than equalizers with OFDM within a wide range of Doppler frequencies. Although the OTFS full MMSE has better performance than the proposed methods at very high Doppler frequency domain, its complexity will be more intensive in terms of hardware implementation due to the banded matrix inversion required. The complexity of the proposed methods (FDE,FDE+DDE) is almost identical to the single-tap equalizer for OFDM systems, because the similar idea has been adopted with the derived simplified channel model, in which the reconstruction of the interference will be much simpler.

\begin{table}[!htbp]
 \begin{center}
     \begin{tabular}{| l | l | }
     \hline
     Parameters & Value \\
     \hline
     Carrier frequency & $5.8$GHz  \\
     \hline
      Bandwidth & $40$MHz \\
     \hline
      Subcarrier spacing  & $78.125$KHz   \\
     \hline
     cyclic-prefix length & $5.12\mu$s   \\
     \hline
     Frame duration for OTFS modulation &$102.4\mu$s\\
     \hline
     Modulation& QPSK\\
     \hline
     Channel model & TU$6$ \cite{goldsmith2005wireless}\\
     \hline
     Channel delay spread& $5\mu$s\\
     \hline

     \end{tabular}
 \end{center}
 \caption{Simulation Parameters}
 \label{Tab:Sim}
 \end{table}


\section{Conclusion}
This paper has derived a equivalent channel matrix of OTFS modulation, which can simplify the OTFS modulation in the transmitter and interference reconstruction in the receiver. With the equivalent channel matrix, a simple two-stage equalizer is proposed to eliminate the impacts of delay spread and Doppler spread for OTFS modulation. It was demonstrated that OTFS modulation with single-tap equalizer can still achieve very promising performance in high Doppler spread scenarios. The interference reconstruction will be much simpler with the equivalent channel matrix.

\ifCLASSOPTIONcaptionsoff
  \newpage
\fi

\bibliographystyle{IEEEtran}
\bibliography{ref_commletter}

\end{document}